\documentclass[12pt,preprintnumbers,nofootinbib,letterpaper]{article}
\pdfoutput=1
\usepackage{amsmath,amssymb,amsfonts,cite,xcolor,epsf,epsfig,epstopdf,graphics,graphicx,mathtools,subcaption,physics,verbatim}
\def\thefootnote{*\arabic{footnote}}
\definecolor{ultramarine}{rgb}{0.07, 0.04, 0.56}
\definecolor{cadmiumgreen}{rgb}{0.0, 0.42, 0.24}
\definecolor{indigo(dye)}{rgb}{0.0, 0.25, 0.42}
\usepackage[linktocpage=true,breaklinks]{hyperref}
\hypersetup{
	colorlinks=true,
	citecolor=ultramarine,
	linkcolor=cadmiumgreen,
	urlcolor=indigo(dye),
}

\usepackage[utf8]{inputenc}

\numberwithin{equation}{section}

\usepackage{array}
\newcolumntype{P}[1]{>{\centering\arraybackslash}p{#1}}

\usepackage[shortlabels]{enumitem}

\usepackage{dcolumn}
\newcolumntype{M}[1]{>{\centering\arraybackslash}m{#1}}
\newcolumntype{N}{@{}m{0pt}@{}}

\usepackage{amsmath}
\usepackage{ulem}

\newcommand{\zbar}{{\bar z}}

\newcommand{\be}{\begin{equation}}  
\newcommand{\ee}{\end{equation}}

\usepackage{tikz}
\begin{document}


\begin{center}

\def\thefootnote{\fnsymbol{footnote}}

\vspace*{1.5cm}
{\Large {\bf Disformal gravitational waves}}
\\[1cm]

{Jibril Ben Achour$^{1,2,3}$, Mohammad Ali Gorji$^{4,5}$, Hugo Roussille$^{3}$}
\\[.7cm]

{\small \textit{$^1$Arnold Sommerfeld Center for Theoretical Physics, Munich, Germany
}}\\

{\small \textit{$^2$Munich Center for Quantum Science and Technology, Germany
}}\\

{\small \textit{$^{3}$Univ de Lyon, ENS de Lyon, Laboratoire de Physique, CNRS UMR 5672, Lyon 69007, France}}\\

{\small \textit{$^{4}$Cosmology, Gravity, and Astroparticle Physics Group, Center for Theoretical Physics of the Universe, Institute for Basic Science (IBS), Daejeon, 34126, Korea
}}\\

{\small \textit{$^{5}$Departament de F\'{i}sica Qu\`{a}ntica i Astrof\'{i}sica, Facultat de F\'{i}sica, Universitat de Barcelona, Mart\'{i} i Franqu\`{e}s 1, 08028 Barcelona, Spain
}}\\

\end{center}

\vspace{1.5cm}

\hrule \vspace{0.5cm}

\begin{abstract}
Contrary to conformal transformations, disformal transformations can change the principal null directions of a spacetime geometry. Thus, depending on the frame a gravitational wave (GW) detector minimally couples to, the properties of GWs may change under a disformal transformation. In this paper, we provide \textit{necessary} and \textit{sufficient} conditions which determine whether GWs change under disformal transformations or not. Our argument is coordinate-independent and can be applied to any spacetime geometry at the fully non-linear level. As an example, we show that an exact radiative solution of massless Einstein-scalar gravity which admits only shear-free parallel transported frame is mapped to a disformed geometry which does not possess any shear-free parallel transported frame. This radiative geometry and its disformed counterpart provide a concrete example of the possibility to generate tensorial GWs from a disformal transformation at the fully non-linear level. This type of non-linear effect can be completely overlooked in the usual linear perturbation theory.
\end{abstract}
\vspace{0.5cm} 

\hrule
\def\thefootnote{\arabic{footnote}}
\setcounter{footnote}{0}

\thispagestyle{empty}


\newpage

\section{Introduction}\label{introduction}
The discovery of late time cosmic acceleration has spurred an exploration of theories beyond general relativity (GR). By far, the most studied theories are scalar-tensor theories where one assumes the presence of a scalar field which characterizes the breathing mode of gravity. A well-known example leading to a new phenomenology is the Brans-Dicke theory. In the absence of matter, this theory is nothing but the conformal field redefinition
\begin{align}\nonumber
g_{\mu\nu} \rightarrow \tilde{g}_{\mu\nu} = C(\phi) g_{\mu\nu} \,,
\end{align}
of the Einstein-scalar system. Even more interestingly, one can realize healthy higher-order scalar-tensor theories by considering a more subtle field redefinition, dubbed disformal transformation (DT) \cite{Bekenstein:1992pj,Kaloper:2003yf, Zumalacarregui:2010wj,Bettoni:2013diz, Zumalacarregui:2013pma,Deruelle:2014zza, Sakstein:2014isa,Brax:2014vva, BenAchour:2016cay}
\begin{align}\label{DT-0}
g_{\mu\nu} \xrightarrow{\mbox{DT}} \tilde{g}_{\mu\nu} = C(\phi, X) g_{\mu\nu} + B(\phi, X) \phi_{\mu} \phi_{\nu} \,,
\end{align}
where $\phi_\mu \equiv \nabla_\mu\phi$ and $X\equiv{g}^{\mu\nu}\phi_\mu\phi_\nu$  is the kinetic energy of the scalar field. Clearly, applying \eqref{DT-0} to any scalar-tensor theory, no new physics will show up as far as the transformation is invertible (see \cite{Deruelle:2014zza,BenAchour:2016cay,Gorji:2018okn,Firouzjahi:2018xob,Jirousek:2018ago,Gorji:2019ttx,Gorji:2020ten,Hammer:2020dqp,Jirousek:2022rym,Jirousek:2022kli,Babichev:2024eoh} for the non-invertible case). However, one can still use invertible DTs to construct new theories by assuming different matter coupling for $g_{\mu\nu}$ and ${\tilde g}_{\mu\nu}$ (see also \cite{Jirousek:2022jhh}). For example, the theory in which matter minimally couples to $g_{\mu\nu}$ is clearly different than the theory in which matter minimally couples to ${\tilde g}_{\mu\nu}$ \cite{Zumalacarregui:2012us,Zumalacarregui:2013pma,vandeBruck:2013yxa,Sakstein:2015jca}. In that case, a natural question is how the observables associated to a given gravitational system are modified when going from the Einstein frame to the disformed (Jordan) frame. This question has been investigated for particular systems in different ways. In cosmology, it turned out that a DT simply induces a change of time coordinate \cite{Domenech:2015hka} and, therefore, all observables will remain unaffected \cite{Minamitsuji:2014waa, Motohashi:2015pra,Cai:2023ykr}. This result allows to capture the cosmological predictions of a large family of modified theories by studying one representant of its disformal class. Yet, for compact objects, a DT is known to introduce new physics \cite{Domenech:2019syf, BenAchour:2020wiw, BenAchour:2020fgy, Anson:2020trg}. Indeed it changes the causal structure by modifying the lightcone direction while the principal null directions (PNDs) of a seed geometry are stretched in the direction of the gradient of the scalar field. Consequently, the Petrov type of a given gravitational field is not stable under a DT in general \cite{Achour:2021pla}.

Thus, while the investigations at the cosmological level show that some key observables are indeed disformally invariant, the non-invariance of the algebraic type of a given metric certainly shows that properties of the gravitational waves (GWs) can change under DTs. \textit{Indeed, any observables that require to specify implicitly how matter couples to the metric, or involve a freely falling observer, can be modified under a DT} \cite{BenAchour:2020wiw, BenAchour:2020fgy, Anson:2020trg}. For radiative geometries which carry non-linear waves, the PNDs and their multiplicity (i.e. their Petrov type) characterize the type of corresponding waves, i.e. breathing, transverse, etc. Therefore, the possibility to deform the PNDs by a DT raises the following question: under which conditions do properties of GWs change after a DT? The goal of this paper is to systematically determine these conditions in a fully non-linear and coordinate-independent manner which can be applied to any spacetime geometry.

To proceed, we investigate how the properties of a null congruence of geodesics are modified under a DT. Concretely, we shall consider a null parallel transported frame (PTF) which realizes the construction of null Fermi coordinates in the neighborhood of a given reference null geodesic \cite{Guedens:2012sz}. Within this specific frame, which can be thought as an idealized GW detector, the behavior of the bundle of light rays (expansion, shear, twist) is encoded in the twelve complex Newman-Penrose spin coefficients \cite{Newman:1961qr}. As expected, a PTF constructed w.r.t the disformed geometry may not retain properties of the seed PTF. The detail of the modifications depends on the profile of the seed scalar field. We confirm this result by applying our general setup to an exact solution of the massless Einstein-scalar system describing the emission of breathing waves by a scalar pulse~\cite{Tahamtan:2015sra}. We explicitly show that while this GR solution admits only shear-free PTFs, the corresponding disformed geometry admits only PTFs with non-vanishing shear. This new exact solution of Horndeski gravity~\cite{BenAchour:2024zzk}, which inherits a non-linear superposition of breathing waves and tensorial GWs, provides an illustrative example for the generation of (tensorial) disformal GWs.

\section{Parallel transported frame}
Consider a spacetime geometry with metric $g_{\mu\nu}$. To analyze the effects of a general DT on this geometry, and in particular on the properties measured by an idealized GW detector, it is convenient to use the local analysis based on the properties of geodesic null congruence (e.g a bundle of light rays) \cite{Achour:2021pla}. We thus introduce a set of tetrads $\theta^{\mu}{}_a$ at any point of the spacetime which allows us to project the metric $g_{\mu\nu}$ into the local rest frame of the observer as
\begin{align}\label{g-Tetrads}
&{ g}_{\mu\nu} = \eta_{ab}\,{\theta }^a{}_{\mu} { \theta}^b{}_{\nu} \,,
&&
\eta_{ab} = { g}_{\mu\nu}{ \theta}^{\mu}{}_a \, { \theta}^{\nu}{}_b \,,
\end{align}
where $a,b=0,1,2,3$ are the Lorentz indices and $\eta_{ab}$ is the Minkowski metric. ${\theta }^a{}_{\mu}$ is the inverse of $\theta^{\mu}{}_a$ such that
\begin{align}\label{Tetrads-complete}
&{ \theta}^a{}_{\mu} {\theta}^{\mu}{}_b = \delta^a{}_b \,, &&
{ \theta}^{\mu}{}_a { \theta}^a{}_{\nu} = \delta^\mu{}_\nu \,.
\end{align}
In order to work with the Newman-Penrose formalism \cite{Newman:1961qr}, we set a null tetrad basis which is given in terms of four complex null vectors
\begin{align}\label{Tetrads}
&{ \theta}^a{}_{\mu} = \big( - { n}_\mu, - { \ell}_\mu, {\bar m}_\mu, { m}_\mu \big) \,, 
&&
{ \theta}^{\mu}{}_a = \big( { \ell}^\mu, { n}^\mu, {m}^\mu, {\bar m}^\mu \big) \,,
\end{align}
where ${ m}^\mu$ and ${ {\bar m}}^\mu$ are complex conjugate to each other, and the Minkowski metric takes the following null form
\begin{eqnarray}\label{eta}
\eta_{ab} = \eta^{ab} \doteq \begin{pmatrix}
0 & -1 & 0 & 0 \\
-1 & 0 & 0 & 0 \\
0 & 0 & 0 & 1 \\
0 & 0 & 1 & 0 
\end{pmatrix} \,.
\end{eqnarray}
 In this null basis, the metric $g_{\mu\nu}$ can be expressed as
\begin{equation}\label{g}
{ g}_{\mu\nu} = - { \ell}_\mu { n}_\nu - { n}_\mu { \ell}_\nu
+ { m}_\mu {\bar { m}}_\nu + {\bar { m}}_\mu { m}_\nu \,,
\end{equation}
where the null vectors satisfy the normalization and orthogonality conditions
\begin{align}\label{null}
&{ g}_{\mu\nu}{ \ell}^\mu{ n}^\nu = -1 \,,
&&{ g}_{\mu\nu}{ m}^\mu{\bar{ m}}^\nu = 1 \,,
\end{align}
and all other contractions between null vectors vanish.

We consider $\ell$ to be geodesic, i.e.
\begin{align}\label{geodesic-l}
D\ell^{\mu} = 0 \,,
\end{align}
where $D \equiv \ell^{\alpha} \nabla_{\alpha}$ is the covariant derivative along $\ell$. The properties of the null congruence are characterized by the twelve complex Newman-Penrose spin coefficients
$\kappa, \epsilon, \pi, \alpha, \beta, \rho, \sigma, \lambda, \nu, \tau, \mu, \gamma$ which are defined in appendix \ref{app-spin-coefficients}. They describe how the bundle of light rays expand, accelerate, shear and twist along each of the null directions ${ \theta}^{\mu}{}_a = \big( { \ell}^\mu, { n}^\mu, {m}^\mu, {\bar m}^\mu \big)$. The vector $\ell$ being geodesic \eqref{geodesic-l}, a PTF can be constructed by further demanding that
\begin{align}\label{PTF}
&D n^\mu = 0 \,, 
&&D m^\mu = 0 \,, 
&&D \bar{m}^\mu = 0 \,,
\end{align}
which in terms of the spin coefficients takes the form
\begin{align}\label{PT-conditions}
&\kappa = -m^{\alpha}D\ell_{\alpha} = 0 \,, 
&& \epsilon = \frac{1}{2}(\bar{m}^{\alpha} Dm_{\alpha} - n^{\alpha}D\ell_{\alpha}) = 0 \,, 
&&\pi = \bar{m}^{\alpha}Dn_{\alpha} = 0 \,.
\end{align}
The tetrads are defined up to Lorentz transformations and the  six degrees of freedom in the Lorentz transformations make it possible to always achieve the above six conditions.

With the PTF \eqref{PT-conditions} at hand, one effectively realizes a set of null Fermi coordinates \cite{Guedens:2012sz}. The remaining spin coefficients associated to this PTF provide the physical quantities characterizing the bundle of light rays that a freely falling observer following a light-like trajectory will measure.

\section{Disformal transformation}
Now, let us investigate how null tetrads change under a DT
\begin{align}\label{Tetrads-dis}
{ \theta}^a{}_{\mu} = \big( -n_\mu , - { \ell}_\mu, {\bar m}^\mu, { m}^\mu \big)
&\xrightarrow{\mbox{DT}} 
{\tilde \theta}^a{}_{\mu}
= \big(- {\tilde n}_\mu,- {\tilde \ell}_\mu, \tilde{{\bar m}}^\mu, \tilde{ m}^\mu \big) \,.
\end{align}
For the sake of simplicity, we restrict our analysis to a pure constant DT Eq. \eqref{DT-0} with 
\begin{align}\label{DT}
&C(\phi, X) =1 \,,
&&B(\phi, X) = B_0 \,,
\end{align}
where $B_0$ is a constant. The equivalence principle\footnote{Throughout this paper, by ``equivalence principle", we refer specifically to the weak equivalence principle \cite{DiCasola:2013iia}.} guaranties that there should exist a set of disformed null tetrads ${\tilde \theta}^{\mu}{}_a$ such that
\begin{align}\label{g-T-Tetrads}
&{\tilde g}_{\mu\nu} = \eta_{ab}\,{\tilde \theta }^a{}_{\mu} {\tilde \theta}^b{}_{\nu} \,,
&&
\eta_{ab} = {\tilde g}_{\mu\nu}{\tilde \theta}^{\mu}{}_a \, {\tilde \theta}^{\nu}{}_b \,.
\end{align}
This tetrad ${\tilde \theta}^{\mu}{}_a$ provides at each point of the geometry the projection onto the local rest frame where the equivalence principle is realized. 
It can be easily shown that under the DT~\eqref{DT}, the relation between the disformed tetrads and original tetrads is given by  \cite{Achour:2021pla}
\begin{align}\label{Tetrads-disform}
{ \theta}^a{}_\mu&\xrightarrow{\mbox{DT}} {\tilde \theta}^a{}_\mu
= J^a{}_b \, { \theta}^b{}_\mu \,,
\end{align}
where
\begin{align}\label{J-def}
&J^a{}_b = \delta^a{}_b + {\cal B}\, \phi^a \phi_b \,;
&&
{\cal B}(X) \equiv \frac{1}{X}\big[\sqrt{1+B_0 X}-1\big] \,,
\end{align}
in which 
\begin{align}
&\phi_a \equiv \phi_{\alpha} \theta^\alpha{}_a \,,
&&\phi^a \equiv g^{\alpha\beta}\phi_{\alpha} \theta^a{}_\beta \,,
\end{align}
such that $\phi_\mu = \phi_a \theta^a{}_\mu$ and $\phi^\mu = \phi^a \theta^\mu{}_a$. Substituting \eqref{J-def} in \eqref{Tetrads-disform} we find that the DT can be written at the level of the tetrad as follows:
\begin{align}\label{Tetrads-disform-explicit}
{\tilde \theta}^a{}_\mu
= { \theta}^a{}_\mu + {\cal B}\, \phi^a \phi_{\mu} \,.
\end{align}
The vector $\phi_\mu$ can be expressed in terms of the null basis as
\begin{align}
\phi_\mu = - \phi_n \ell_\mu - \phi_\ell n_\mu 
+ \phi_{\bar m} m_\mu + \phi_m {\bar m}_\mu \,.
\end{align}
Substituting this in \eqref{Tetrads-disform-explicit}, we find the explicit expression of the disformed null vectors in terms of the seed null vectors.

Equipped with the disformed null vectors, it is straightforward to compute the disformed spin coefficients. In particular, we find
\begin{align}\label{PTF-dis}
\begin{split}
\kappa &\xrightarrow{\mbox{DT}} \tilde{\kappa} = \kappa + \kappa_{\rm DT} \,,
\\
\epsilon &\xrightarrow{\mbox{DT}} \tilde{\epsilon} = \epsilon + \epsilon_{\rm DT} \,,  
\\
\pi &\xrightarrow{\mbox{DT}} \tilde{\pi} = \pi + \pi_{\rm DT} \,,
\end{split} 
\end{align}
such that $\kappa_{\rm DT}=\epsilon_{\rm DT}=\pi_{\rm DT}=0$ for $B_0=0$ (see appendix \ref{app-spin-coefficients} for the explicit forms of $\kappa_{\rm DT}, \epsilon_{\rm DT}, \pi_{\rm DT}$ to first order in $B_0$). The results \eqref{PTF-dis} clearly show that, in general, the PTF \eqref{PT-conditions} does not remain a PTF after performing the DT since 
\begin{align}\label{PT-conditions-DT}
&\tilde{\kappa} \neq 0 \,, 
&&\tilde{\epsilon} \neq 0 \,, 
&&\tilde{\pi} \neq 0 \,,
\end{align} 
as  in general $\kappa_{\rm DT}\neq0$, $\epsilon_{\rm DT}\neq0$, $\pi_{\rm DT}\neq0$.
In particular, it shows that the DT induces a deviation w.r.t the geodesic motion (see Fig.~\ref{Fig}). Let us further expand on this point.

\section{Lorentz transformations and matter coupling} 

As we shall see, it is instructive to contemplate the effects of the DT on the local rest frames $\tilde{\eta}_{ab}$ (resp. $\eta_{ab}$) associated to $\tilde{g}_{\mu\nu}$ (resp. $g_{\mu\nu}$). Indeed, this reveals that when one chooses a local rest frame in which the equivalence principle is realized, one implicitly fixes the matter coupling since matter must be minimally coupled to the metric associated to this local rest frame.

To see this, recall that the local Minkowski metric is invariant under the Lorentz transformations (LTs) 
\begin{align}
\eta_{ab} \Lambda^a{}_{c} \Lambda^b{}_{d} = \eta_{cd} \,,
\end{align}
where $\Lambda^a{}_b$ is any LT (which is characterized by six parameters). Substituting the above relation in \eqref{g-Tetrads} and \eqref{g-T-Tetrads}, we see that both seed tetrads $\theta^a{}_\mu$ and disformed tetrads ${\tilde \theta}^a{}_\mu$ are defined up to Lorentz transformations (LTs). 

Let us rewrite the first equation in \eqref{g-T-Tetrads} as
\begin{align}\label{g-T-Tetrads-LT}
&{\tilde g}_{\mu\nu} 
= \eta_{ab}\,{\tilde \theta }^a{}_{\mu} {\tilde \theta}^b{}_{\nu} 
= \eta_{ab} J^a{}_{c} J^b{}_{d} \, { \theta }^c{}_{\mu} { \theta}^d{}_{\nu}  \,,
\end{align}
Comparing the above result with \eqref{g-Tetrads}, we see that the disformed metric does not take the local Minkowski form when we express it in terms of the original tetrads since, in general, one has 
\begin{align}
\eta_{ab} J^a{}_{c} J^b{}_{d} \neq \eta_{cd} \,.
\end{align}
Now the key point is that the generator $J^a{}_{b}$ is not a Lorentz transformation. Since this $J$-map encodes the effects of the disforml transformation at the level of the local rest frame, it implies that the effects of a disformal transformation cannot be compensated by a Lorentz transformation. Indeed, using the definition \eqref{J-def}, one can compare the two generators which read
\begin{align}
J_{ab} & = \eta_{ac} J^c{}_b = \eta_{ab} + {\cal B}\, \phi_a \phi_b \\
\Lambda_{ab} & = \eta_{ac} \Lambda^c{}_b = \eta_{ab} + \omega_{ab} \qquad \text{with} \qquad \omega_{ab}=-\omega_{ba}
\end{align} 
Thus, the antisymmetric part $\omega_{ab}$ cannot compensate the effects of the symmetric disformed part $\phi_a\phi_b = \phi_b \phi_a$.

The appearance of 
\be
\eta_{ab} J^a{}_{c} J^b{}_{d} \neq \eta_{cd}
\ee 
in \eqref{g-T-Tetrads-LT} shows that if the equivalence principle is considered w.r.t. $g_{\mu\nu}$, a free fall observer in $g_{\mu\nu}$ is not a free fall observer in ${\tilde g}_{\mu\nu}$. This is because the two tetrads/observers ${ \theta}^a{}_{\mu}$ and ${\tilde \theta}^a{}_{\mu}$ are not related to each other through LTs $\Lambda^a{}_b$ but through the $J$-map $J^a{}_b$ defined in \eqref{Tetrads-disform}. The matter minimally couples to the metric w.r.t which the equivalence principle is realized, i.e. which can take locally the Minkowski form. \textit{Thus, the choice of tetrad w.r.t which the equivalence principle is realized implicitly fixes the coupling of matter fields.} The matter minimally couples to $g_{\mu\nu}$ (resp. to ${\tilde g}_{\mu\nu}$) if one works with the tetrad ${\theta}^a{}_{\mu}$ (resp. ${\tilde \theta}^a{}_{\mu}$). \textit{Hence, depending on the tetrad one works with, one deals with a different theory: by choosing tetrads, one implicitly fixes the matter coupling.} 

In particular, that is why the Petrov type changes under a DT \cite{Achour:2021pla}. From this discussion, the $J$-map ${\tilde \theta}^a{}_\mu = J^a{}_b \, { \theta}^b{}_\mu$, that is defined in Eq. \eqref{Tetrads-disform}, can be understood as the operation encoding the departure from a local Minkowski frame in which the equivalence principle is realized, i.e. the effective distortion of the local freely falling frame induced by the DT. Obviously, one has always the choice to decide to assign the free fall and thus the realization of the equivalence principle to ${\tilde \theta}^a{}_\mu$ or to ${ \theta}^b{}_\mu$.

We have shown that starting from PTF \eqref{PT-conditions} and performing the DT, we end up with \eqref{PT-conditions-DT} which is not PTF. We, however, still have the freedom to perform LTs and bring \eqref{PT-conditions-DT} to PTF. We thus perform LTs on the disformed null vectors $\theta^\mu{}_a=(\tilde{\ell}^\mu, \tilde{n}^\mu, \tilde{m}^\mu, \bar{\tilde{m}}^\mu)$. The general LTs of the null tetrad basis are given by \cite{Chandrasekhar:1985kt}
\begin{align}\label{LTs}
\begin{split}
\tilde{\ell}^\mu &\xrightarrow{\mbox{LT}} \tilde{\ell}'^\mu = \frac{1}{A} \tilde{\ell}^\mu + A |{\rm b}|^2 \tilde{n}^\mu 
+ 2 {\rm Re}\big[ e^{i \theta } \bar{\rm b} \tilde{m}^\mu \big]  
\,, 
\\
\tilde{n}^\mu &\xrightarrow{\mbox{LT}} \tilde{n}'^\mu 
= \frac{1}{A} |{\rm a}|^2 \tilde{\ell}^\mu
+ A |1+{\rm a}\bar{\rm b}|^2 \tilde{n}^\mu + 2 {\rm Re}\big[ e^{i \theta } \bar{\rm a} \left( 1 + {\rm a} \bar{\rm b} \right) \tilde{m}^\mu \big] 
\,, 
\\ 
\tilde{m}^\mu &\xrightarrow{\mbox{LT}} \tilde{m}'^\mu =  \frac{1}{A} {\rm a} \tilde{\ell}^\mu
+ A {\rm b} \left(1 + {\rm a} \bar{\rm b} \right) \tilde{n}^\mu + 2 {\rm a} {\rm Re}\big[ e^{i \theta } \bar{\rm b} \tilde{m}^\mu \big] + e^{i \theta } \tilde{m}^\mu \,, 
\end{split}
\end{align}
where the functions $(a,b)$ are free complex functions which characterize class I and II transformations while $A$ and $\theta$ are real functions that characterize class III transformation \cite{Chandrasekhar:1985kt}. These free functions correspond to the six free parameters of the local Lorentz group.

After performing general LTs on the disformed null basis $\theta^\mu{}_a=(\tilde{\ell}^\mu, \tilde{n}^\mu, \tilde{m}^\mu, \bar{\tilde{m}}^\mu)$, the disformed spin coefficients \eqref{PTF-dis} transform as 
\begin{align}\label{SC-LT}
\begin{split}
\kappa &\xrightarrow{\mbox{DT}} \tilde{\kappa} 
\xrightarrow{\mbox{LT}} \tilde{\kappa}' 
= \kappa + \kappa_{\rm DT} + \kappa_{\rm LT}(A,\theta,{\rm a},{\rm b})
\,, 
\\
\epsilon &\xrightarrow{\mbox{DT}} \tilde{\epsilon}
\xrightarrow{\mbox{LT}} \tilde{\epsilon}' 
= \epsilon + \epsilon_{\rm DT} + \epsilon_{\rm LT}(A,\theta,{\rm a},{\rm b})
\,,
\\
\pi &\xrightarrow{\mbox{DT}} \tilde{\pi}
\xrightarrow{\mbox{LT}} \tilde{\pi}' 
= \pi + \pi_{\rm DT} + \pi_{\rm LT}(A,\theta,{\rm a},{\rm b})
\,.
\end{split}
\end{align} 
The explicit forms of $\kappa_{\rm LT}, \epsilon_{\rm LT}, \pi_{\rm LT}$ to first order in $B_0$ are shown in appendix \ref{app-LT}. Of course, one can compute them at the fully nonlinear level. However, they take a rather complicated form which we will not present here.

So far, we have clarified how a DT changes the properties of a seed PTF, i.e. a parallel transported null frame constructed w.r.t the seed geometry $g_{\mu\nu}$. As we have shown explicitly, the PTF condition is not stable under a (constant) DT.
The natural question at this stage is: what happens if one now constructs a PTF w.r.t the disformed metric $\tilde{g}_{\mu\nu}$? For a radiative spacetime, what are the general conditions for the properties of GWs (which are encoded in the spin coefficients associated to a PTF) to be modified under a DT? We answer this question in the next section.

\section{Disformed parallel transported frame}

With the Lorentz transformed null basis \eqref{SC-LT} at hand, we now construct the {\it disformed} PTF (i.e. w.r.t the disformed metric $\tilde{g}_{\mu\nu}$), imposing that
\begin{align}\label{PT-conditions-LT}
&\tilde{\kappa}' = 0 \,, 
&&\tilde{\epsilon}' = 0 \,, 
&&\tilde{\pi}' = 0 \,.
\end{align} 
Starting from the PTF seed \eqref{PT-conditions}, conditions \eqref{PT-conditions-LT} imply
\begin{align}\label{PT-conditions-LT-kappa-epsilon-pi}
\kappa_{\rm LT} &= - \kappa_{\rm DT} \,, &&\epsilon_{\rm LT}= - \epsilon_{\rm DT} \,, &&\pi_{\rm LT}= - \pi_{\rm DT} \,.
\end{align}
The six free functions in $a, b, A, \theta$ can be chosen to always satisfy the above conditions. The {\it necessary} condition to change the properties of GWs under the DT is that at least one of the free functions $a, b, A, \theta$ acquires a non-vanishing value after imposing \eqref{PT-conditions-LT}.
\begin{figure}
	\centering
	\includegraphics{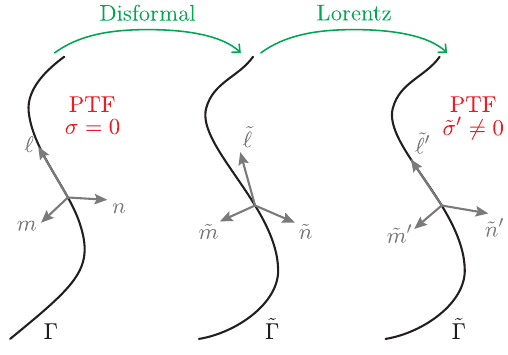}
	\caption{Transformation of a PTF of the seed under i) a disformal transformation and ii) a sequence of Lorentz transformations to build a disformed PTF. This procedure does not leave the spin coefficients invariant, e.g., shear-free PTF can turn into a PTF with non-vanishing shear.}\label{Fig}
\end{figure}

With the disformed PTF \eqref{PT-conditions-LT} at hand, one can compute the remaining disformed spin coefficients. Let us focus on the shear which transforms as
\begin{align} \label{sigma-DT-LT}
\sigma \xrightarrow{\mbox{DT}} \tilde{\sigma}
\xrightarrow{\mbox{LT}}
\tilde{\sigma}' 
& = \sigma + \sigma_{\rm DT} + \sigma_{\rm LT}
\,.
\end{align}
where we have separated the effects induced by the DT and the effects of the LT.
Having already used the six free functions in the LTs to construct the PTF~\eqref{PT-conditions-LT}, in general, one has $\sigma_{\rm DT} \neq - \sigma_{\rm LT}$ which implies that
\begin{equation}
\tilde{\sigma}' \neq \sigma \,.
\label{eq:cond-sigma-order}
\end{equation}
The same argument applies to the remaining eight spin coefficients such that, in general we have
\begin{align}\label{eq:cond-SC-order}
\tilde{\alpha}'\neq\alpha,
\quad 
\tilde{\beta}'\neq\beta,
\quad 
\tilde{\rho}'\neq\rho,
\quad
\tilde{\lambda}'\neq\lambda,
\quad 
\tilde{\nu}'\neq\nu, 
\quad
\tilde{\tau}'\neq\tau, 
\quad
\tilde{\mu}'\neq\mu,
\quad
 \tilde{\gamma}'\neq\gamma \,.
\end{align}
In this regard, the {\it sufficient} condition to change the properties of GWs under the DT is that at least one of the conditions in Eqs.~\eqref{eq:cond-sigma-order} and \eqref{eq:cond-SC-order} satisfies.

The key point we wish to emphasize here is that, in full generality, for a radiative seed, a DT will modify the properties of the GWs. In particular, even if the seed PTF is shear-free $\sigma=0$, one may find $\tilde{\sigma}'\neq0$ w.r.t the disformed PTF (see Fig.~\ref{Fig}). Thus, while a free falling (lightlike) observer w.r.t. the seed geometry can conclude on the absence of shear, i.e. $\sigma =0$, a free falling observer w.r.t. the disformed geometry can detect non-vanishing shear, i.e. $\tilde{\sigma}' \neq 0$. In the next section, we will explicitly confirm this fact by considering a concrete example.

\section{A concrete example} 
To provide a concrete example, we need to pick up a seed geometry. We consider an exact radiative solution of the Einstein-scalar system presented in~\cite{Tahamtan:2015sra}. The metric is given by
\begin{align}\label{met}
&\dd{s}^2 = - K(z,\zbar) \dd w^2 - 2 \dd w \dd r + \frac{r^2 - \chi^2(w)}{P^2(z,\zbar)} \dd{z}\dd{\zbar} \,, 
\end{align}
while the scalar profile is given by a pure time-dependent monopole
\begin{align}
\label{scalprof}
&\phi(w,r) = \frac{1}{\sqrt{2}} \log{\left( \frac{r - \chi}{r + \chi}\right)} \,.
\end{align}
This geometry describes the gravitational radiation emitted by a scalar pulse\footnote{As shown in \cite{Tahamtan:2015sra}, this solution of the canonical Einstein-Scalar system is radiative, i.e. it exhibits a non-vanishing flux of energy at $\mathcal{I}^{+}$. The algebraic classification performed in \cite{Podolsky:2016sff} shows that the solution is either of Petrov type D when $K$ is a constant, or of Petrov type II otherwise.}. The pulse is encoded in the function $\chi(w)$ that vanishes in both remote past and future, when $w\rightarrow \pm w_0$, and reaches its maximum $\chi(0) = \chi_\mathrm{max}$ when $w=0$~\cite{BenAchour:2024zzk}. Here, $\chi_\mathrm{max}$ plays the role of the scalar charge. As we are only interested in using this seed geometry to illustrate our general argument, we do not provide any detail on the properties of this geometry. The interested reader is referred to \cite{Tahamtan:2015sra, Podolsky:2016sff, BenAchour:2024zzk} for further details.

To analyze this solution, we consider the following set of null vectors
\begin{align}\nonumber
&\ell^{\mu} = \delta^\mu_r \,,
&&n^{\mu} = \delta^\mu_w - \frac{1}{2} K \delta^\mu_r \,,
&&m^{\mu} = \frac{-iP}{\sqrt{2(r^2 - \chi^2)}}\, \delta^\mu_{\bar{z}} \,,
\end{align}
which satisfy the desired normalization conditions \eqref{null} and form a null basis for the seed metric~\eqref{met}. Computing the spin coefficients for this tetrad, one finds that the only non-vanishing spin coefficients are
\begin{align}
\label{Seed-SC-RT}
&\beta = -\bar{\alpha} 
= \frac{i \partial_{\zbar} P}{\sqrt{2} \sqrt{r^2-\chi ^2}} \,,
&&
\mu = - \frac{K \rho +2 \chi  \chi '}{2\left(r^2-\chi ^2\right)} \,, 
&&\rho = -\frac{r }{r^2-\chi ^2} \,, 
&&
\nu = \frac{i P \partial_zK}{\sqrt{2} \sqrt{r^2-\chi ^2}} \,.
\end{align}
We see that the seed satisfies the PTF conditions \eqref{PT-conditions}. Moreover, the seed is shear-free 
\begin{align}\label{RT-shear-0}
\sigma=0 \,,
\end{align}
and also twist-free $\omega=-\Im(\rho)=0$. Indeed, in Ref. \cite{Tahamtan:2015sra} it is proved that this radiative geometry carries pure breathing waves encoded in the non-vanishing expansion $\Theta = - {\rm Re}(\rho)\neq0$. See \cite{Eardley:1973zuo}
 for a discussion on the different possible polarizations.
 
The disformed metric reads
\begin{align}
\dd{s}^2 & = - K(z,\bar{z})  \dd{w}^2 - 2 \dd{w} \dd{r} + \frac{r^2 - \chi^2(w)}{P^2(z,\zbar)}\dd{z}\dd{\zbar}  \\
& \;\;\; \;+ B_0 \left[ \phi^2_w  \dd{w}^2 + 2 \phi_w \phi_{r} \dd{w}\dd{r} + \phi^2_{r} \dd{r}^2 \right] \,,
\end{align}
where  $\phi_w = \partial_w \phi$ and $\phi_r = \partial_r \phi$. The above metric provides an exact radiative solution of a shift-symmetric subset of the Horndeski theory. A detailed analysis of this new solution has been presented in \cite{BenAchour:2024zzk}. Following the discussion above, we emphasize again that while one could misleadingly see this as a pure field redefinition of the GR solution, the analysis of the spin coefficients associated to a PTF constructed w.r.t to the GR or the Horndeski solution reveals that it is not the case. The fact that the field redefinition introduced new physics is precisely due to the fact that analyzing the GWs content of the seed and its disformed version requires to fix a freely falling frame w.r.t one or the other geometries, implicitly fixing in that way to which metric the matter is minimally coupled. This automatically implies that one is dealing with two inequivalent theories. Let us now apply the previous construction explicitly. 

Here we are interested in analyzing the properties of a PTF in this new geometry following the approach described in the previous sections. Solving the PTF condition 
\eqref{PT-conditions-LT-kappa-epsilon-pi} is straightforward but difficult. We thus work in the regime of $B_0\ll1$ and we show that the shear is non-vanishing in the disformed frame at the second order in $B_0$. This is enough for our purpose since it proves that the vanishing shear $\sigma=0$ in \eqref{RT-shear-0} becomes non-vanishing in the disformed frame, i.e. $\tilde{\sigma}'\neq0$. This is an explicit realization of the general result \eqref{eq:cond-sigma-order}. Note that one can always implement our general setup to perform the full calculation without assuming $B_0\ll1$.

To first order in $B_0$, the PTF condition for the disformed frame~\eqref{PT-conditions-LT-kappa-epsilon-pi} gives
\begin{align}
- D {\rm b}^{(1)} + {\rm b}^{(1)} \rho &= 0 \,, \label{eq:PT-kappa-1}
\end{align}
while the shear is given by
\begin{align}
\tilde{\sigma}' & = - B_0 \Big( \delta {\rm b}^{(1)} - 2 {\rm b}^{(1)} {\beta} \Big) \,. \label{eq:PT-sigma-1}
\end{align}
Using $\beta =-(\delta\ln{P})/2$ and substituting $D {\rm b}^{(1)}$ from~\eqref{eq:PT-kappa-1}, one finds 
\begin{align}
D\ln\tilde{\sigma}' = \rho \,.
\end{align}
The above result follows from 
\begin{align}
D\delta\ln{\rm b}^{(1)}=\delta{D}\ln{\rm b}^{(1)}=\delta\rho=0
\end{align}
since $\rho$ depends only on $r$ and $w$. This shows that the shear satisfies the same equation as ${\rm b}^{(1)}$ in \eqref{eq:PT-kappa-1} such that $\tilde{\sigma}' \propto {\rm b}^{(1)}$. Integrating~\eqref{eq:PT-kappa-1} gives
\begin{align}
{\rm b}^{(1)} = \frac{f}{\sqrt{r^2 - \chi^2}}
\end{align}
where $f(w, z, \zbar)$ is a complex free function that characterizes the residual gauge freedom in  class II LTs after having imposed the PTF conditions. Choosing the gauge $f=0$ gives ${\rm b}^{(1)}=0$ and we find
\begin{align}
&\tilde{\sigma}' \propto B_0 {\rm b}^{(1)} = 0 \,.
\end{align}
Thus, at first order in $B_0$, LTs of class I and III are enough to obtain a disformed PTF and, more importantly, shear can be set to zero. Therefore, we have to go to the next order in $B_0$.

At second order in $B_0$, the PTF condition~\eqref{PT-conditions-LT-kappa-epsilon-pi}  reads
\begin{align}\label{PT-kappa-2}
&- D{\rm b}^{(2)} + {\rm b}^{(2)} \rho = \frac{1}{4} \bar{\nu} \phi_\ell^4 \,,
&&\phi_{\ell}=\phi_r \,,
\end{align}
where we have imposed ${\rm b}^{(1)}=0$. The shear becomes
\begin{align}
&\tilde{\sigma}' = - B_0^2 \big( \delta {\rm b}^{(2)} - 2 {\rm b}^{(2)} \beta \big) \,. \label{eq:PT-sigma-2} 
\end{align}
It is then straightforward to find
\begin{align}
&D\tilde{\sigma}' = - B_0^2 D{\rm b}^{(2)} \delta \ln\big(P D{\rm b}^{(2)} \big) \neq  0 \,,
\end{align}
where we have used the fact Eq.  that\eqref{PT-kappa-2} implies
\begin{align}
{\rm b}^{(2)}\neq0  \qquad \text{since} \qquad \phi_\ell\neq0
\end{align}
\textit{Therefore, the shear \eqref{eq:PT-sigma-2} associated to the disformed PTF is non-vanishing at the second order in $B_0$. It follows that contrary to the seed,  the disformed solution can no longer exhibit any shear-free PTF.} One can integrate~\eqref{PT-kappa-2} and substitute the result in~\eqref{eq:PT-sigma-2} to find an explicit solution for the shear. See Ref. \cite{BenAchour:2024zzk} for details. However, the fact that the shear is non-vanishing in any PTF of the disformed solution is enough to demonstrate our argument. 

How one should understand this new property? The presence of this non-vanishing shear associated to any PTF shows that the new disformed radiative solution now carries also source for the tensorial GWs on top of the breathing ones already present in the seed. The rather surprising point is that the scalar field profile is a simple time-dependent monopole in both cases. The generation of the physical shear (as measured by a light-like observer sitting in any given PTF) can be understood as originating from the higher-order terms in the dynamics which transforms the scalar monopole into  higher-order multipoles \cite{BenAchour:2024zzk}. Thus, higher-order scalar-tensor theories have the ability to provide a new source of tensorial GWs, even in the presence of a pure (time-dependent) scalar monopole.

\section{Summary}
Depending on the frame GW detectors minimally couple to, properties of GWs can change after performing a DT. Focusing on spin coefficients, which characterize the behavior of a bundle of light rays as seen by a geodesic lightlike observer, we have shown that, in general, a DT maps a PTF to a non-PTF (see Fig.~\ref{Fig}). Then performing LTs, we brought the disformed non-PTF to the PTF form. Based on this systematic method, we have found the {\it necessary} and {\it sufficient} conditions under which the properties of GWs change through a DT. Our results are coordinate-independent and can be applied to any radiative spacetime geometry at the fully non-linear level. 

To clarify the importance of this general result, we have applied our setup to a particular example, belonging to the Robinson-Trautman geometries for massless Einstein-scalar gravity, which is shear-free. By definition, this solution always admits a shear-free PTF and the only GWs carried by this geometry are breathing \cite{Podolsky:2016sff}. The corresponding disformed geometry was analyzed in detail in \cite{BenAchour:2024zzk}. Here, we demonstrate that in the disformed geometry, it is impossible to construct a PTF which is shear-free. It shows that this non-linear radiative solution acquires a non-vanishing physical shear, such that the DT sources tensorial GWs on top of the breathing ones. It provides a concrete example of the generation of disformal tensorial GWs at the fully non-linear level. This type of non-linear effect can be completely overlooked in the usual linear perturbations theory.

\vspace{0.7cm}

{\bf Acknowledgments:} The work of MAG was supported by IBS under the project code IBS-R018-D3 and Mar\'{i}a Zambrano fellowship.

\vspace{0.7cm}

\appendix

\section{Spin coefficients}\label{app-spin-coefficients}

The twelve complex Newman-Penrose spin coefficients for the null tetrad $\theta^\mu{}_a=(\ell^\mu, n^\mu, m^\mu, \bar{m}^\mu)$ are given by \cite{Chandrasekhar:1985kt}
\begin{align}\nonumber
&\kappa \equiv -m^{\alpha}D\ell_{\alpha} \,,
&&
\epsilon \equiv \frac{1}{2}(\bar{m}^{\alpha} Dm_{\alpha} - n^{\alpha}D\ell_{\alpha}) \,,
&&
\pi \equiv \bar{m}^{\alpha}Dn_{\alpha} \,,
\\ \nonumber
&\sigma \equiv -m^{\alpha}\delta\ell_{\alpha} \,,
&&
\beta \equiv \frac{1}{2}(\bar{m}^{\alpha}\delta{m}_{\alpha}
- n^{\alpha}\delta\ell_{\alpha}) \,,
&&
\mu \equiv \bar{m}^{\alpha}\delta {n}_{\alpha} \,,
\\ \nonumber
&\rho \equiv -m^{\alpha}\bar{\delta}\ell_{\alpha} \,,
&&
\alpha \equiv \frac{1}{2}(\bar{m}^{\alpha}\bar{\delta}m_{\alpha} - n^{\alpha}\bar{\delta}\ell_{\alpha}) \,,
&&
\lambda \equiv \bar{m}^{\alpha}\bar{\delta}n_{\alpha} \,,
\\
&\tau \equiv -m^{\alpha}\Delta\ell_{\alpha} \,,
&&
\gamma \equiv \frac{1}{2}(\bar{m}^{\alpha} \Delta m_{\alpha}
-n^{\alpha}\Delta\ell_{\alpha}) \,,
&&
\nu \equiv \bar{m}^{\alpha}\Delta n_{\alpha} \,, \nonumber
\end{align}
where the directional derivatives along the null vectors are defined as
\begin{align}\nonumber
&D = \ell^\alpha \nabla_\alpha \,,
&&
\Delta = n^\alpha \nabla_\alpha \,,
&&
\delta = m^\alpha \nabla_\alpha \,,
&&
\bar{\delta} = \bar{m}^\alpha \nabla_\alpha \,.
\end{align}
The usual kinematical quantities like the expansion $\Theta$, shear $\sigma_s$, and twist (vorticity) $\omega$ can be deduced from the spin coefficients. For the null vector $\ell^\mu$, they are defined as
\begin{align}
&\Theta \equiv - {\rm Re}\left[\rho\right] \,, 
&&\sigma_s \equiv |\sigma| \,,
&&\omega \equiv - {\rm Im}\left[\rho\right] \,.
\end{align}

Having the disformed null vectors \eqref{Tetrads-disform} in hand, one can easily compute the disformed spin coefficients which are defined in the spirit of the disformed null basis $(\tilde{\ell}, \tilde{n}, \tilde{m}, \bar{\tilde{m}})$. One then finds
\begin{align}\nonumber
\kappa_{\rm DT} 
&= \tilde{\kappa} - \kappa =  -\tilde{m}^{\mu}\tilde{D}\tilde{\ell}_{\mu} + {m}^{\mu}{D}{\ell}_{\mu} = B_0 \kappa_{\rm DT}^{(1)} + \cdots \,, 
\\
\nonumber
\epsilon_{\rm DT} &= \tilde{\epsilon} - \epsilon 
= \frac{1}{2} (\bar{\tilde{m}}^{\mu} \tilde{D}\tilde{m}_{\mu} - \tilde{n}^{\mu}\tilde{D}\tilde{\ell}_{\mu}) 
- \frac{1}{2} (\bar{{m}}^{\mu} {D}{m}_{\mu} - {n}^{\mu}{D}{\ell}_{\mu}) 
= \epsilon + B_0 \epsilon_{\rm DT}^{(1)} + \cdots \,, \hspace{-.1cm}
\\ \nonumber
\pi_{\rm DT} &= \tilde{\pi} - \pi 
= \bar{\tilde{m}}^{\mu}\tilde{D}\tilde{n}_{\mu} 
- \bar{{m}}^{\mu}{D}{n}_{\mu} 
= \pi + B_0 \pi_{\rm DT}^{(1)} + \cdots \,,
\end{align}
where
\begin{align}\nonumber
\kappa_{\rm DT}^{(1)} &\equiv 
\kappa \phi _n \phi _\ell 
+\frac{1}{2} (2\bar{\epsilon } -\rho) \phi _m \phi _\ell 
+\frac{1}{2} \left(\tau -\bar{\pi }\right) \phi_\ell^2
- {\rm Re}\left[\bar{\kappa } \phi_m \right] \phi_m 
+ \frac{1}{2} [\phi_\ell,\phi_m]_D \,,
\\
\epsilon_{\rm DT}^{(1)} &\equiv
\frac{1}{2} \big[ \left(\bar{\epsilon }+2 \epsilon \right) \phi _n
-\alpha  \phi_m -\left(\bar{\pi }+\beta \right) \phi_{\bar{m}} \big] \phi _\ell 
- \frac{1}{2}\bar{\kappa }  \phi _m\phi _n
\nonumber \\ \nonumber
&+{\rm Re}\left[\gamma\right] \phi _\ell^2 
- i {\rm Im}\left[\epsilon\right] |\phi_{m}|^2 
+ \frac{1}{4} \big( [\phi_\ell,\phi_n]_D
- [\phi_m,\phi_{\bar{m}}]_D \big)
\,,
\\ \nonumber
\pi_{\rm DT}^{(1)} &\equiv 
-\frac{1}{2} \left(\mu  \phi_{\bar{m}}+\lambda  \phi _m-2 \pi  \phi_n\right) \phi _\ell
+  \bar{\epsilon } \phi _n \phi _{\bar{m}} 
-\frac{1}{2} \bar{\kappa }
\phi _n^2 
+\frac{1}{2} \nu  \phi _\ell^2
- {\rm Re}\left[\pi  \phi _m\right] 
- \frac{1}{2} [\phi_n,\phi_{\bar{m}}]_D
\,, \hspace{2.7cm}
\end{align}
where 
\begin{align}
[\phi_\ell,\phi_m]_D\equiv\phi_\ell D\phi_m - \phi_m D \phi_\ell \,.
\end{align}
In the last steps, for the matter of presentation, we have expanded the results for $B_0\ll1$ such that $\cdots$ denotes the terms that are quadratic and higher in $B_0$. We emphasize that one can compute all $\kappa_{\rm DT}$, $\epsilon_{\rm DT}$, $\pi_{\rm DT}$ at the fully non-linear order in $B_0$.

\section{Lorentz transformations up to first order in $B_0$}\label{app-LT}
The general LTs of the null tetrad basis are given by \eqref{LTs}. Considering parametrization 
\begin{align}\nonumber
&{\rm a} =B_0 {\rm a}^{(1)} + \cdots \,, 
&&{\rm b}=B_0 {\rm b}^{(1)}+ \cdots \,, 
&&A = 1 +B_0 A^{(1)}+ \cdots \,,
&&\theta=B_0\theta^{(1)}+ \cdots \,,
\end{align}
such that the LTs reduce to the unity map for $B_0=0$, \eqref{LTs} simplify to
\begin{align}\label{LTs-1stOrder}
\begin{split}
\tilde{\ell}'^\mu &= \tilde{\ell}^\mu - B_0 \big( A^{(1)} {\ell}^\mu - 2 {\rm Re}\big[{\rm b}^{(1)}\bar{m}^\mu\big] \big) + \cdots 
\,, 
\\
\tilde{n}'^\mu 
& = \tilde{n}^\mu + B_0 \big(A^{(1)} n^\mu
+ 2 {\rm Re}\big[{\rm a}^{(1)}\bar{m}^\mu\big]\big) + \cdots 
\,, 
\\ 
\tilde{m}'^\mu 
& = \tilde{m}^\mu + B_0 \big({\rm a}^{(1)} \ell^{\mu} + {\rm b}^{(1)} n^{\mu} + i \theta^{(1)} m^\mu\big) + \cdots 
\,.
\end{split}
\end{align} 

It is then straightforward to find
\begin{align}
\kappa_{\rm LT} &= \tilde{\kappa}' - \tilde{\kappa}
= B_0 \kappa_{\rm LT}^{(1)} + \cdots 
\,, 
\nonumber \\ \nonumber 
\epsilon_{\rm LT} &= \tilde{\epsilon} - \tilde{\epsilon} = B_0 \epsilon_{\rm LT}^{(1)} + \cdots  
\,, 
\\ \nonumber
\pi_{\rm LT} &= \tilde{\pi}' - \tilde{\pi} 
= B_0 \pi_{\rm LT}^{(1)} + \cdots 
\,, 
\end{align} 
where  
\begin{align*}\nonumber
\kappa_{\rm LT}^{(1)} &\equiv 
\big(i \theta^{(1)} -2 A^{(1)} \big) \kappa
+ {\rm b}^{(1)} ( \rho +2 \epsilon )
+ \bar{\rm b}^{(1)} \sigma
- Db^{(1)}
\,,
\\
\pi_{\rm LT}^{(1)} &\equiv
2 \bar{\rm a}^{(1)} \epsilon + \bar{\rm b}^{(1)}  \mu + {\rm b}^{(1)} \lambda - i \theta^{(1)}  \pi + D\bar{\rm a}^{(1)} 
\,,\\
\epsilon_{\rm LT}^{(1)} &\equiv \!\begin{aligned}[t]
&\bar{\rm a}^{(1)} \kappa + \bar{\rm b}^{(1)} \beta - A^{(1)} \epsilon + {\rm b}^{(1)} (\alpha+\pi) - \frac{1}{2} \left(D(A^{(1)} + i \theta^{(1)})\right) \,. \end{aligned}
\end{align*}

\bibliographystyle{JHEPmod}
\bibliography{refs}

\end{document}